\documentclass[12pt]{article}

\newcommand{\be}{\begin{equation}}
\newcommand{\ee}{\end{equation}}
\usepackage{amsmath}
\usepackage{amssymb}
\usepackage{latexsym}
\usepackage{epsfig}

\newcommand{\bea}{\begin{eqnarray}}
\newcommand{\eea}{\end{eqnarray}}

\newcommand{\ti}{\tilde}

\newcommand{\RRR}{{\hbox{\rm R\kern-2.35mm R}}}

\def\ZZZ{{\hbox{ Z\kern-1.6mm Z}}}

\def\d{\delta}

\def\L{\Lambda}

\begin{document}

\begin{titlepage}
\rightline{December 2015}
\rightline{  Imperial-TP-2015-CH-01}
\begin{center}
\vskip 2.5cm
{\Large \bf {
Finite   Gauge Transformations and Geometry in Extended Field Theory}}\\
\vskip 2.0cm
{\large {N.  Chaemjumrus and C.M. Hull  }}
\vskip 0.5cm
{\it {The Blackett Laboratory}}\\
{\it {Imperial College London}}\\
{\it {Prince Consort Road}}\\
{\it { London SW7 2AZ, U.K.}}\\

\vskip 2.5cm
{\bf Abstract}
\end{center}

\vskip 0.5cm

\noindent
\begin{narrower}
The recently derived expressions for finite gauge transformations in double field theory with duality group $O(d,d)$ are generalised to give 
expressions for finite gauge transformations for extended field theories with duality group
$SL(5,\mathbb {R})$, $SO(5,5)$
and $E_6$. The generalised metrics are discussed.

\end{narrower}

\end{titlepage}

\newpage

\tableofcontents
\baselineskip=16pt
\section{Introduction}

 String theory on a toroidal background  gives rise to a double field theory, with  fields  depending on 
a doubled spacetime in which  the   periodic coordinates $x^i$ on the torus are supplemented by 
 dual periodic coordinates $\ti x_i$ conjugate to the  winding numbers. Key features of double field theory (DFT) are that T-duality is manifest and the fields depend on all the doubled coordinates. The double field theory corresponding to the metric, $b$-field and dilaton of the bosonic string   was derived from string theory in \cite{Hull:2009mi} to cubic order in the fields.
The full theory has proved rather intractable, and much work has been done on a small subsector of the theory, obtained by imposing the \lq strong constraint', which locally implies that locally all fields and parameters depend on only half the doubled coordinates. The strongly constrained theory has been found to all orders in the fields \cite{Hohm:2010jy,Hohm:2010pp}, and
 is locally equivalent to the conventional field theory of metric, $b$-field and dilaton, and   DFT reduces to the duality-covariant formulation of field theory proposed by Siegel \cite{Siegel:1993th}, and can be thought of as a formulation  in terms of generalised geometry 
  \cite{Hitchin:2004ut} -  \cite{Lee:2015xga}. 
 Further details on the history of DFT relevant here are given in \cite{Hull:2014mxa}
 and reviews and further references are given in 
  \cite{Aldazabal:2013sca,Berman:2013eva, Hohm:2013bwa}.

There has been much work on extending this to formulations of supergravity in which U-duality is manifest.
Generalised geometry has a natural action of the continuous group $O(d,d)$ and in
\cite{Hull:2007zu, Pacheco:2008ps}
   this was generalised to extended geometries in which there is a natural action of a duality group $E_d$. 
   Doubled geometry  with extra coordinates conjugate to string winding modes then has a natural generalisation to an extended geometry  with extra coordinates conjugate to brane wrapping modes \cite{Hull:2004in}, with a natural action of   $E_d$ duality transformations.
   Extended field theories (EFT) generalise strongly constrained DFT to a theory on an extended geometry that is covariant under $E_d$ U-duality transformations \cite{Hull:2004in}, 
   \cite{Berman:2010is} - \cite{Bossard:2015foa}.     See \cite{Berman:2013eva}
 for a review  and further references on EFT.

DFT was derived for toroidal backgrounds, but the resulting theory has a certian background independence \cite{Hohm:2010pp}, suggesting that DFT could be apllicable to more general backgrounds. To address such questions requires a better understanding of transition functions and global structure, and for this one needs formulae for gauge transformations with finite parameters.
The infinitesimal gauge transformations of DFT or EFT are given by the action of the so-called generalised Lie derivative. As this is given by the usual Lie derivative in the doubled or extended spacetime, it is tempting to think of these in terms of some generalisation of a diffeomorphism in the doubled or extended spacetime.
However, the gauge algebra is very different from that of diffeomorphisms in the extended space, and is in fact locally equivalent to the algebra of diffeomorphisms and $p$-form gauge transformations in the original (unextended) spacetime. There have been a number of  papers seeking expressions for finite gauge transformations in DFT 
\cite{Hohm:2013bwa}, \cite{Hohm:2012gk} - \cite{Papadopoulos:2014ifa}.
There are a number of problems with some of these proposals, as has been discussed in \cite{Papadopoulos:2014mxa} and  \cite{Hull:2014mxa}. In particular, the proposal of \cite{Hohm:2012gk}
 attempts to realise the gauge transformations as diffeomorphisms in the doubled space, with novel transformation properties for generalised tensors. This runs into difficulties as the gauge algebra is not that of diffeomorphisms  \cite{Hull:2014mxa}, and realising $b$-field transformations as diffeomorphisms of   dual coordinates is problematic in general \cite{Papadopoulos:2014mxa}.

In  \cite{Hull:2014mxa}, explicit forms were found for finite gauge transformations in DFT that have the correct gauge algebra, are in agreement with the finite transformations for the metric and $b$-field, and which make explicit contact with generalised geometry. The purpose of this paper is to extend these results to extended field theories with duality group $E_d$. We will consider the cases
$E_4=
SL(5,\mathbb {R})$, $E_5=SO(5,5)$
and $E_6$; we expect similar results will hold for $E_7$.
While this paper was in preparation, the paper \cite{Rey:2015mba}
 appeared addressing the same question, but using an approach which appears to suffer from the same issues as the approach of \cite{Hohm:2012gk}.

\section{Finite Transformations for Double Field Theory}\setcounter{equation}{0}


In this section, we review the results of \cite{Hull:2014mxa}.
Double field theory has  fields  on a doubled space-time $  M$ with coordinates $X^M$,
where the indices $M,N=1,\dots 2D$ transform covariantly under the action of $O(D,D)$.
Indices are  raised and lowered
using the constant $O(D,D)$ invariant matrix 
  $\eta_{MN}$ of the form
  \be
\label{etais}
\eta_{MN} =  \begin{pmatrix}
0&1 \\1&0 \end{pmatrix}\,.
\ee
The fields and  gauge parameters   satisfy 
 the   `strong constraint'  (so that $\partial^M\partial_M A = 0$  
 and $\partial^M A\, \partial_M B=0$
for any fields or parameters $A$ and $B$).
It was shown in  \cite{Hohm:2010jy}  
 that the strong constraint implies that,
at least locally, all fields are restricted to a $D$-dimensional   subspace  that is null with respect to  
 $\eta$.

Consider   a patch $\cal U$ of $  M$ with coordinates 
\be
X^M = \begin{pmatrix}   x^m \\ \tilde x_m \end{pmatrix} \, 
\ee
 where $m=1,\dots,D$.
A generalised vector $W^M$  transforms as a vector under $O(D,D)$ and decomposes as 
\be
W^M = \begin{pmatrix}   w^m \\ \tilde w_m \end{pmatrix} \, ,
\ee
under $GL(D,\mathbb{R})$.
  The strong constraint is solved by having
 all fields independent of $\tilde x_m$ so that
\be
 \tilde \partial^m  =0
 \ee
 on all fields and parameters.
 Then the fields just depend on the coordinates $x^m$, parameterising a $D$-dimensional patch $U$ (which can be thought of as the quotient of ${\cal U}$ by the isometries generated by
 $\partial/\partial \ti x^m$).

The generalised Lie derivative
\be
\hat{\mathcal{L}}_V W^M = V^P\partial_P W^M+W^P(\partial^MV_P-\partial_P V^M)
\ee
for $V^M(x), W^M(x)$ then gives
\be
(\hat{\mathcal{L}}_V W)^m = v^p\partial_p w^m-w^p\partial_pv^m = \mathcal{L}_v w^m
 \ee
 and
\begin{eqnarray}
(\widehat{{\cal L}}_{V} W)_{m} \   &=& \ v^{p}\partial_{p} \tilde w_{m}{}+ 
  \tilde w_{p}\partial_m v^{p}
  +w^{p} (\partial_{m}\tilde v_{p} 
  \, - \partial_{p}\tilde v_m)\,
  \\ &=& \ 
   {\cal L}_v
      \tilde w  _{m}{}
  +w^{p} (\partial_{m}\tilde v_{p} 
  \, - \partial_{p}\tilde v_m)
\end{eqnarray}
where $ {\cal L}_v  
$  is the usual Lie derivative on $U$.

Under an infinitesimal transformation with parameter $V^M$,  $W^M$ transforms as 
\be
\delta W^M=  \widehat{{\cal L}}_{V} W^{M} 
\ee
giving
\begin{eqnarray}
\delta w^m &=&{\cal L}_v w^m
\\
\delta \tilde w_m &=& {\cal L}_v
      \tilde w  _{m}{}
  +w^{p} (\partial_{m}\tilde v_{p} 
  \, - \partial_{p}\tilde v_m)
\end{eqnarray}
It will be convenient to rewrite the components of the generalised vector $W$ as
\be
w = w^m e_m  \qquad \tilde{w}_{(1)} = \tilde{w}_m e^m,
\ee
where 
$e_m= \partial/\partial x^m$ and $e^m=dx^m$ are the coordinate bases 
for $TU$ and $T^*U$, respectively. Then the generalised vector can be written as
\be
W = w\oplus\tilde{w}_{(1)}.
\ee
Under an infinitesimal transformation with parameter $V$, these transform as 
\begin{eqnarray}
\delta_Vw &=& \mathcal{L}_v w,\\
\delta_V\tilde{w}_{(1)} &=& \mathcal{L}_v \tilde{w}_{(1)}-\iota_wd\tilde{v}_{(1)},
\end{eqnarray}
where $\mathcal{L}_v$ is a Lie derivative on patch $U$.

Next, we
introduce a gerbe connection $B_{(2)}$ on $U$,
\be
B_{(2)} = \frac{1}{2}B_{mn}e^m \wedge e^n,
\ee 
which transforms under the gauge transformation as
\begin{eqnarray}
\delta_{V}B_{(2)} = \mathcal{L}_v B_{(2)}+d\tilde{v}_{(1)}.
\end{eqnarray}
Then  \begin{eqnarray}
\hat{w}_{(1)} &=& \tilde{w}_{(1)}+\iota_wB_{(2)},
\end{eqnarray}
  transforms as
\begin{eqnarray}
\delta_V \hat{w}_{(1)} &=& \mathcal{L}_v \hat{w}_{(1)}.
\end{eqnarray}
and so is a 1-form on $U$, and is invariant under the $\tilde v$ transformations.
Then
\be
\hat W = w\oplus \hat{w}_{(1)}
\ee
is a section of $(T\oplus T^*)U$.

The finite transformation of $\hat{W}$ is given by
\begin{eqnarray}
{w}'(x') = {w}(x)  \qquad
\hat{w}_{(1)}'(x') = \hat{w}_{(1)}(x),
\end{eqnarray}
where $x'(x) = e^{-v^m\partial_m}x$.
Using the finite transformation of the coordinate bases  
\be
e'_m(x') = e_n(x)\frac{\partial x^n}{\partial x'^m}  \qquad e'^m(x') = e^n(x)\frac{\partial x'^m}{\partial x^n},
\ee
the finite transformation of the components of $\hat{W}$ are then
\be \label{hattrans}
{w}'^m(x') = w^n(x)\frac{\partial x'^m}{\partial x^n}  \qquad \hat{w}'_{m}(x') = \hat{w}_{n}(x)\frac{\partial x^n}{\partial x'^m}.
\ee

The finite transformation of the gerbe connection can be taken to be
\begin{eqnarray}
\label{BTranss}
B_{(2)}'(x') = B_{(2)}(x)+d\tilde{v}_{(1)}(x),
\end{eqnarray}
so that 
\begin{eqnarray}
\hat{w}_{(1)}'(x') = \tilde{w}_{(1)}'(x')+\iota_{w'}B_{(2)}'(x').
\end{eqnarray}
This then gives the finite transformations of 
$\tilde{w}_{(1)}$:
\begin{eqnarray}
\tilde{w}_{(1)}'(x') &=& \hat{w}_{(1)}'(x')-\iota_{w'}B'_{(2)}(x'),\nonumber\\
&=& \hat{w}_{(1)}(x)-\iota_w(B_{(2)}(x)+d\tilde{v}_{(1)}(x)),\nonumber\\
&=& \tilde{w}_{(1)}(x)-\iota_wd\tilde{v}_{(1)}(x).
\end{eqnarray}

To summarise,  the transformation of $W$ is given by
\begin{eqnarray}
w'(x') &=& w(x),\\
\tilde{w}_{(1)}'(x') &=& \tilde{w}_{(1)}(x)-\iota_wd\tilde{v}_{(1)}(x).
\end{eqnarray}
which implies   the components
$(w^m, \tilde w_m)$ transform as
\begin{eqnarray}
w'^m(x') &=& w^n(x)\frac{\partial x'^m}{\partial x^n}\\
\tilde{w}'_{m}(x') &=& \left[\tilde{w}_{p}(x)-w^n(x)(\partial_n \tilde{v}_p(x) -\partial_p \tilde{v}_n(x))\right]\frac{\partial x^p}{\partial x'^m}.
\end{eqnarray}
\section{Finite 
Transformations for Extended Field Theory} \setcounter{equation}{0}
\subsection{$SL(5,\mathbb{R}$) Extended field theory}

In $SL(5,\mathbb{R})$ extended field theory  \cite{Berman:2011cg,Blair:2014zba}, a generalised vector $W^M$ transforms as a {\bf 10} of $SL(5,\mathbb{R})$ 
where  the indices $M,N =1, \dots, 10$ label the {\bf 10} representation of $SL(5,\mathbb{R})$. 
It decomposes under $GL (4,\mathbb{R})\subset SL(5,\mathbb{R})$ into a vector and 2-form:
\be
W^M = \begin{pmatrix}   w^m \\ \tilde{w}_{mn}\end{pmatrix} \, ,
\ee
where
$m,n=1, \dots, 4$  
and $\tilde{w}_{mn}=-\tilde{w}_{nm}$.
The coordinates in a patch $\mathcal{U}$ consist of 7 spacetime coordinates $y^\mu$, $\mu =0,\dots , 6$, together with 10 internal coordinates $X^M$  transforming as a {\bf 10} of $SL(5,\mathbb{R})$.
This decomposes under $GL (4,\mathbb{R})\subset SL(5,\mathbb{R})$ as 
\be
X^M = \begin{pmatrix}   x^m \\ \tilde{x}_{mn}\end{pmatrix} \,,
\ee
where, in a suitable duality frame,  $x^m$ are the usual coordinates  on $T^4$ and $\tilde{x}_{mn}$ are periodic coordinates conjugate to M2-brane wrapping numbers on $T^4$.
Fields and gauge parameters depend on both $y^\mu $ and $X^M$, but we will suppress dependence on $y^\mu$ in what follows.

The strong constraint of SL($5,\mathbb{R})$ EFT is given by
\be
\epsilon^{iMN}\epsilon_{iPQ} \partial_M (\dots) \partial_N(\dots) = 0 ,
\ee
where $(\dots)$ represents fields and gauge parameters, and the indices $i,j = 1,\dots,5$ label the fundamental representation $\bf{5}$ of $SL(5,\mathbb{R})$. An index $M$ can be regarded as an antisymmetric pair of indices $[m_1m_2]$, so that 
 $\epsilon^{iMN} = \epsilon^{im_1 m_2 n_1n_2}$.
 The strong constraint can be solved such that the fields are independent of wrapping coordinates $\tilde{x}_{mn}$ so that
 \be
 \tilde{\partial}^{mn}(\dots) = 0,
 \ee
 where $\tilde{\partial}^{mn} = \frac{\partial}{\partial \tilde{x}_{mn}}$.
The gauge transformations of SL$(5,\mathbb{R})$ extended field theory are given by the generalised Lie derivative, which is defined as
\begin{eqnarray}
\hat{\mathcal{L}}_{V}W^M = V^N\partial_NW^M-W^N\partial_NV^M+\epsilon^{iMN}\epsilon_{iPQ}\partial_{N}V^{P}W^{Q}.
\end{eqnarray}

It is convenient to rewrite the components of the generalised vector as
\begin{eqnarray}
w &=& w^m e_m,\\
\tilde{w}_{(2)} &=& \frac{1}{2!}\tilde{w}_{mn}e^m\wedge e^n.
\end{eqnarray} Then the generalised vector $W$ is
\be
W = w\oplus \tilde{w}_{(2)}.
\ee
Under a gauge transformation with  gauge parameter $V$, a generalised vector $W$ transforms as
\be
\delta_V W = \hat{\mathcal{L}}_V W \, .
\ee
This decomposes into 
\begin{eqnarray}
\delta_V w &=& \mathcal{L}_vw,\\
\delta_V \tilde{w}_{(2)} &=& \mathcal{L}_v\tilde{w}_{(2)}-\iota_wd\tilde{v}_{(2)},
\end{eqnarray}
where $\mathcal{L}_v$ is the ordinary Lie derivative with a parameter $v$.

Next we introduce a gerbe connection $C_{(3)}$, 
\be
C_{(3)} = \frac{1}{3!}C_{mnp}e^m\wedge e^n \wedge e^p,
\ee
which transforms under a gauge transformation as
\begin{eqnarray}
\delta_V C_{(3)} = \mathcal{L}_vC_{(3)}+d\tilde{v}_{(2)} \, .
\end{eqnarray}
This allows us to define $\hat{w}_{(2)}$ by
\begin{eqnarray}
\hat{w}_{(2)} &=& \tilde{w}_{(2)}+\iota_wC_{(3)}.
\end{eqnarray}
Under a gauge transformation, this transforms as a 2-form
\begin{eqnarray}
\delta_V\hat{w}_{(2)} &=& \mathcal{L}_v \hat{w}_{(2)}
\end{eqnarray} 
and is invariant under the $\tilde v$ transformations.
Therefore, $\hat{W} = {w} \oplus \hat{w}_{(2)}$ is a section of $(T\oplus \Lambda^2 T^*) U$. 
This allows us to immediately write down the 
 finite transformation of $\hat{W}$, which is   given by
\begin{eqnarray}
 {w}'(x') &=&  {w}(x),\\
\hat{w}_{(2)}'(x') &=& \hat{w}_{(2)}(x),
\end{eqnarray}
where $x' = e^{-v^m\partial_m}x$.

Using the finite transformation of the coordinate bases  given by
\be
e'_m(x') = e_n(x)\frac{\partial x^n}{\partial x'^m}  \qquad e'^m(x') = e^n(x)\frac{\partial x'^m}{\partial x^n},
\ee
the finite transformation of the components of $\hat W$ are then 
\begin{eqnarray}
 {w}'^m(x') &=& w^n(x)\frac{\partial x'^m}{\partial x^n},  \\
\hat{w}'_{mn}(x') &=& \hat{w}_{pq}(x)\frac{\partial x^p}{\partial x'^m}\frac{\partial x^q}{\partial x'^n}.
\end{eqnarray}
The finite transformation of the gerbe connection can be taken to be
\begin{eqnarray}
C'_{(3)}(x') = C_{(3)}(x)+d\tilde{v}_{(2)}(x),
\end{eqnarray}
so that
\begin{eqnarray}
\hat{w}_{(2)}'(x') &=& \tilde{w}_{(2)}'(x')+\iota_{w'}C'_{(3)}(x').
\end{eqnarray}
This then gives the finite transformation of $\tilde{w}_{(2)}$:
\begin{eqnarray}
\tilde{w}_{(2)}'(x') &=& \hat{w}'_{(2)}(x')-\iota_{w'}C'_{(3)}(x'),\nonumber\\
&=& \hat{w}_{(2)}(x)-\iota_w(C_{(3)}(x)+d\tilde{v}_{(2)}(x)),\nonumber\\
&=& \hat{w}_{(2)}(x)-\iota_wC_{(3)}(x)-\iota_wd\tilde{v}_{(2)}(x),\nonumber\\
&=& \tilde{w}_{(2)}(x)-\iota_wd\tilde{v}_{(2)}(x).
\end{eqnarray}

To summarise, the transformation of $W$ is given by
\begin{eqnarray}
w'(x') &=& w(x),\\
\tilde{w}_{(2)}'(x') &=& \tilde{w}_{(2)}(x)-\iota_wd\tilde{v}_{(2)}(x),
\end{eqnarray}
which implies the components $(w^m, \tilde{w}_{mn}) $ transform as
\begin{eqnarray}
w'^m(x') &=& w^n(x)\frac{\partial x'^m}{\partial x^n},\\
\tilde{w}'_{mn}(x') &=& \left(\tilde{w}_{pq}(x)-3w^r(x)(d\tilde{v}_{(2)})_{rpq}(x)\right)\frac{\partial x^p}{\partial x'^m} \frac{\partial x^q}{\partial x'^n},
\end{eqnarray}
where $(d\tilde{v}_{(2)})_{rpq} = \partial_{[r}\tilde{v}_{pq]}$.
\subsection{$SO(5,5)$ Extended field theory}

In $SO(5,5)$ extended field theory  \cite{Berman:2011pe,Abzalov:2015ega},
a generalised vector $W^M$ transforms as a spinor  of $SO(5,5)$, where the indices $M, N = 1,\dots 16$ label the positive chirality spinor representation.
It decomposes under $GL (5,\mathbb{R})\subset SO(5,5)$ into
\be
W^M = \begin{pmatrix}   w^m \\ \tilde{w}_{mn} \\ \tilde{w}_{mnpqr} \end{pmatrix} \, ,
\ee
where $m,n = 1,\dots,5$, $\tilde{w}_{mn} = -\tilde{w}_{nm}$, and $\tilde{w}_{mnpqr} = \tilde{w}_{[mnpqr]}$.

The coordinates in a patch $\mathcal{U}$ consist of 6 spacetime coordinates $y^\mu$, $\mu =0,\dots , 5$, together with 16 internal coordinates $X^M$  transforming as a {\bf 16} of $SO(5,5)$.
This decomposes under $GL (5,\mathbb{R})\subset SO(5,5)$ into
\be
X^M = \begin{pmatrix}   x^m \\ \tilde{x}_{mn} \\ \tilde{x}_{mnpqr} \end{pmatrix} \,,
\ee
 where, in a suitable duality frame,  $x^m$ are the usual coordinates  on $T^5$ and $\tilde{x}_{mn}$ are periodic coordinates conjugate to M2-brane wrapping numbers on $T^5$
 and $\tilde{x}_{mnpqr}$ are periodic coordinates conjugate to M5-brane wrapping numbers on $T^5$.
Fields and gauge   parameters depend on both $y^\mu $ and $X^M$, but we will suppress dependence on $y^\mu$ in what follows.

The strong constraint of $SO(5,5)$ EFT is given by
\be
\gamma^{MN}_a\gamma^a_{PQ} \partial_M (\dots) \partial_N(\dots) = 0 ,
\ee
where $(\dots)$ represents fields and gauge parameters, $\gamma^a_{MN}$ is a gamma matrix of $SO(5,5)$ and $a=1,\dots 10$ labels the vector representation of $SO(5,5)$.
The strong constraint can be solved such that the fields are independent of wrapping coordinates $\tilde{x}_{mn}$ and $\tilde{x}_{mnpqr}$ so that
 \be
 \tilde{\partial}^{mn}(\dots) = 0 \text{ and } \tilde{\partial}^{mnpqr}(\dots) = 0,
 \ee
 where $\tilde{\partial}^{mn} = \frac{\partial}{\partial \tilde{x}_{mn}}$ and $\tilde{\partial}^{mnpqr} = \frac{\partial}{\partial \tilde{x}_{mnpqr}}.$
The gauge transformation of $SO(5,5)$ extended field theory is given by the generalised Lie derivative, which is  
\be
\hat{\mathcal{L}}_V W^M = V^N\partial_N W^M-W^N\partial_N V^M +\frac{1}{2}\gamma^{MN}_a\gamma^a_{PQ}\partial_N V^P W^Q,
\ee

It is convenient to rewrite the components of the generalised vector as
\begin{eqnarray}
w &=& w^m e_m,\\
\tilde{w}_{(2)} &=& \frac{1}{2!}\tilde{w}_{mn}e^m\wedge e^n,\\
\tilde{w}_{(5)} &=& \frac{1}{5!}\tilde{w}_{mnpqr} e^m\wedge e^n\wedge e^p\wedge e^q\wedge e^r.
\end{eqnarray}Then the generalised vector $W$ is
\be
W = w\oplus \tilde{w}_{(2)}\oplus\tilde{w}_{(5)}.
\ee
Under a gauge transformation with gauge parameter $V$, the generalised vector $W$ transforms as
\be
\delta_V W = \hat{\mathcal{L}}_V W,
\ee
which decomposes into
\begin{eqnarray}
\delta_V w &=& \mathcal{L}_vw,\\
\delta_V \tilde{w}_{(2)} &=& \mathcal{L}_v\tilde{w}_{(2)}-\iota_wd\tilde{v}_{(2)},\\
\delta_V \tilde{w}_{(5)} &=& \mathcal{L}_v\tilde{w}_{(5)}-\tilde{w}_{(2)}\wedge d\tilde{v}_{(2)}.
\end{eqnarray}
where $\mathcal{L}_v$ is an ordinary Lie derivative with a parameter $v$.

Next we define a gerbe connection $C_{(3)}$, 
\be
C_{(3)} = \frac{1}{3!}C_{mnp}e^m\wedge e^n \wedge e^p,
\ee
which transforms under a gauge transformation as
\begin{eqnarray}
\delta_V C_{(3)} = \mathcal{L}_vC_{(3)}+d\tilde{v}_{(2)}.
\end{eqnarray}
This allows us to define $\hat{w}_{(2)}$ and $\hat{w}_{(5)}$ by
\begin{eqnarray}
\hat{w}_{(2)} &=& \tilde{w}_{(2)}+\iota_wC_{(3)},\\
\hat{w}_{(5)} &=& \tilde{w}_{(5)}+\tilde{w}_{(2)}\wedge C_{(3)}+\frac{1}{2}\iota_w C_{(3)}\wedge C_{(3)}.
\end{eqnarray}
Under a gauge transformation, these objects transform as a 2-form and a 5-form, respectively,
\begin{eqnarray}
\delta_V\hat{w}_{(2)} &=& \mathcal{L}_v \hat{w}_{(2)},\\
\delta_V\hat{w}_{(5)} &=& \mathcal{L}_v \hat{w}_{(5)},
\end{eqnarray}
and are invariant under the $\tilde{v}$ transformations. Therefore, $\hat{W} = {w}\oplus \hat{w}_{(2)}\oplus \hat{w}_{(5)}$ is a section of $(T\oplus \Lambda^2 T^*\oplus \Lambda^5 T^*) U$. The finite transformations are then
\begin{eqnarray}
{w}'(x') &=& {w}(x),\\
\hat{w}_{(2)}'(x') &=& \hat{w}_{(2)}(x),\\
\hat{w}_{(5)}'(x') &=& \hat{w}_{(5)}(x),
\end{eqnarray}
where $x' = e^{-v^m\partial_m}x$.

Using the finite transformation of the coordinate bases given by
\be
e'_m(x') = e_n(x)\frac{\partial x^n}{\partial x'^m}  \qquad e'^m(x') = e^n(x)\frac{\partial x'^m}{\partial x^n},
\ee
the finite transformation of the components of $\hat{W}$ are
\begin{eqnarray}
{w}'^m(x') &=& w^n(x)\frac{\partial x'^m}{\partial x^n},  \\
\hat{w}'_{mn}(x') &=& \hat{w}_{mn}(x)\frac{\partial x^p}{\partial x'^m}\frac{\partial x^q}{\partial x'^n},\\
\hat{w}'_{mnpqr}(x') &=& \hat{w}_{stuvw}(x)\frac{\partial x^s}{\partial x'^m} \frac{\partial x^t}{\partial x'^n} \frac{\partial x^u}{\partial x'^p} \frac{\partial x^v}{\partial x'^q} \frac{\partial x^w}{\partial x'^r}
\end{eqnarray}

The finite transformation of the gerbe connection can be taken to be
\begin{eqnarray}
C'_{(3)}(x') = C_{(3)}(x)+d\tilde{v}_{(2)}(x),
\end{eqnarray}
so that
\begin{eqnarray}
\hat{w}_{(2)}'(x') &=& \tilde{w}_{(2)}'(x')+\iota_{w'}C'_{(3)}(x').\\
\hat{w}_{(5)}'(x') &=& \tilde{w}_{(5)}'(x')+\tilde{w}_{(2)}'(x')\wedge C'_{(3)}(x')+\frac{1}{2}\iota_{w'}C'_{(3)}(x')\wedge C'_{(3)}(x').
\end{eqnarray}
This give the finite transformation of $\tilde{w}_{(2)}$:
\begin{eqnarray}
\tilde{w}_{(2)}'(x') &=& \hat{w}'_{(2)}(x')-\iota_{w'}C'_{(3)}(x').\nonumber\\
&=& \hat{w}_{(2)}(x)-\iota_w(C_{(3)}(x)+d\tilde{v}_{(2)}(x)),\nonumber\\
&=& \hat{w}_{(2)}(x)-\iota_wC_{(3)}(x)-\iota_wd\tilde{v}_{(2)}(x),\nonumber\\
&=& \tilde{w}_{(2)}(x)-\iota_wd\tilde{v}_{(2)}(x).
\end{eqnarray}
Furthermore, the finite transformations of $\tilde{w}_{(5)}$ are
\begin{eqnarray}
\tilde{w}_{(5)}'(x') &=& \hat{w}_{(5)}'(x')-\tilde{w}_{(2)}'(x')\wedge C'_{(3)}(x')-\frac{1}{2}\iota_{w'}C'_{(3)}(x')\wedge C'_{(3)}(x'),\nonumber\\
&=& (\tilde{w}_{(5)}(x)+\tilde{w}_{(2)}(x)\wedge C_{(3)}(x)+\frac{1}{2}\iota_w C_{(3)}(x)\wedge C_{(3)}(x))\nonumber\\
& &-(\tilde{w}_{(2)}(x)-\iota_wd\tilde{v}_{(2)}(x))\wedge(C_{(3)}(x)+d\tilde{v}_{(2)}(x))\nonumber\\
& &-\frac{1}{2}\iota_w(C_{(3)}(x)+d\tilde{v}_{(2)}(x))\wedge(C_{(3)}(x)+d\tilde{v}_{(2)}(x)),\nonumber\\
&=& \tilde{w}_{(5)}(x)-\tilde{w}_{(2)}(x)\wedge d\tilde{v}_{(2)}(x)+\frac{1}{2}\iota_wd\tilde{v}_{(2)}(x)\wedge d\tilde{v}_{(2)}(x).
\end{eqnarray}

To summary, the transformation of $W$ is given by
\begin{eqnarray}
w'(x') &=& w(x),\\
\tilde{w}_{(2)}'(x') &=& \tilde{w}_{(2)}(x)-\iota_wd\tilde{v}_{(2)}(x),\\
\tilde{w}_{(5)}'(x') &=& \tilde{w}_{(5)}(x)-\tilde{w}_{(2)}(x)\wedge d\tilde{v}_{(2)}(x)+\frac{1}{2}\iota_wd\tilde{v}_{(2)}(x)\wedge d\tilde{v}_{(2)}(x),
\end{eqnarray}
which implies the components $(w^m, \tilde{w}_{mn}, \tilde{w}_{mnpqr})$ transform as
\begin{eqnarray}
w'^m(x') &=& w^n(x)\frac{\partial x'^m}{\partial x^n},\\
\tilde{w}'_{mn}(x') &=& \left(\tilde{w}_{pq}(x)-3w^r(x)(d\tilde{v}_{(2)})_{rpq}(x)\right)\frac{\partial x^p}{\partial x'^m} \frac{\partial x^q}{\partial x'^n},\\
\tilde{w}'_{mnpqr}(x') &=&\left(\tilde{w}_{stuvw}(x)-30\tilde{w}_{[st}(x)(d\tilde{v}_{(2)})_{uvw]}(x)+15w^l(x)(d\tilde{v}_{(2)})_{l[st}(x)(d\tilde{v}_{(2)})_{uvw]}(x)\right)\nonumber\\
& & \times \frac{\partial x^s}{\partial x'^m} \frac{\partial x^t}{\partial x'^n} \frac{\partial x^u}{\partial x'^p} \frac{\partial x^v}{\partial x'^q} \frac{\partial x^w}{\partial x'^r},
\end{eqnarray}
where $(d\tilde{v}_{(2)})_{rpq} = \partial_{[r}\tilde{v}_{pq]}$.
\subsection{$E_6$ Extended field theory}

In $E_6$ extended field theory  \cite{Hohm:2013vpa,Musaev:2015pla,Baguet:2015xha},
a generalised vector $W^M$ transforms in the fundamental ({\bf 27}) representation  of $E_6$
with    $M,N =1, \dots, 27$.
It decomposes under $GL (6,\mathbb{R})\subset E_6$ into
\be
W^M = \begin{pmatrix}   w^m \\ \tilde{w}_{mn} \\ \tilde{w}_{mnpqr} \end{pmatrix} \, ,
\ee
where $m, n = 1,\dots, 6$, $\tilde{w}_{mn} = -\tilde{w}_{nm}$, and $\tilde{w}_{mnpqr} = \tilde{w}_{[mnpqr]}$.

The coordinates in a patch $\mathcal{U}$ consist of 5 spacetime coordinates $y^\mu$, $\mu =0,\dots , 4$, together with 27 internal coordinates $X^M$  transforming as a {\bf 27} of $E_6$.
This decomposes under $GL (6,\mathbb{R})\subset E_6$ into
\be
X^M = \begin{pmatrix}   x^m \\ \tilde{x}_{mn} \\ \tilde{x}_{mnpqr} \end{pmatrix} \,,
\ee
 where, in a suitable duality frame,  $x^m$ are the usual coordinates  on $T^6$ and $\tilde{x}_{mn}$ are periodic coordinates conjugate to M2-brane wrapping numbers on $T^6$
 and $\tilde{x}_{mnpqr}$ are periodic coordinates conjugate to M5-brane wrapping numbers on $T^6$.
Fields and gauge parameters depend on both $y^\mu $ and $X^M$, but we will suppress dependence on $y^\mu$ in what follows.

The strong constraint of $E_6$ EFT is given by
\be
c^{MNR}c_{PQR} \partial_M (\dots) \partial_N(\dots) = 0 ,
\ee
where $(\dots)$ represents fields and gauge parameters,and $c^{MNP}$ and $c_{MNP}$ are the $E_6$ invariant tensors.
The strong constraint can be solved such that the fields are independent of wrapping coordinates $\tilde{x}_{mn}$ and $\tilde{x}_{mnpqr}$ so that
 \be
 \tilde{\partial}^{mn}(\dots) = 0 \text{ and } \tilde{\partial}^{mnpqr}(\dots) = 0,
 \ee
 where $\tilde{\partial}^{mn} = \frac{\partial}{\partial \tilde{x}_{mn}}$ and $\tilde{\partial}^{mnpqr} = \frac{\partial}{\partial \tilde{x}_{mnpqr}}.$
The gauge transformation of $E_6$ extended field theory is given by the generalised Lie derivative, which is defined as
\begin{eqnarray}
\hat{\mathcal{L}}_V W^M = V^N \partial_N W^M - W^N \partial_N V^M + 10c^{MNR}c_{PQR}\partial_N V^Q W^R,
\end{eqnarray}

It is convenient to rewrite the components of the generalised vector as
\begin{eqnarray}
w &=& w^m e_m,\\
\tilde{w}_{(2)} &=& \frac{1}{2!}\tilde{w}_{mn}e^m\wedge e^n,\\
\tilde{w}_{(5)} &=& \frac{1}{5!}\tilde{w}_{mnpqr} e^m\wedge e^n\wedge e^p\wedge e^q\wedge e^r.
\end{eqnarray}
Then the generalised vector $W$ is
\be
W = w\oplus \tilde{w}_{(2)}\oplus\tilde{w}_{(5)}.
\ee
Under a gauge transformation with a gauge parameter $V$, the generalised vector W transform as
\be
\delta_V W = \hat{\mathcal{L}}_V W,
\ee
which decomposes into
\begin{eqnarray}
\delta_V w &=& \mathcal{L}_vw,\\
\delta_V \tilde{w}_{(2)} &=& \mathcal{L}_v\tilde{w}_{(2)}-\iota_wd\tilde{v}_{(2)},\\
\delta_V \tilde{w}_{(5)} &=& \mathcal{L}_v\tilde{w}_{(5)}-\tilde{w}_{(2)}\wedge d\tilde{v}_{(2)}-\iota_wd\tilde{v}_{(5)}.
\end{eqnarray}
where $\mathcal{L}_v$ is an ordinary Lie derivative with a parameter $v$.

Next we introduce gerbe connections $C_{(3)}$ and $C_{(6)}$,
\begin{eqnarray}
C_{(3)} &=& \frac{1}{3!}C_{mnp}e^m \wedge e^n \wedge e^p,\\
C_{(6)} &=& \frac{1}{6!}C_{mnpqrs} e^m \wedge e^n \wedge e^p \wedge e^q \wedge e^r \wedge e^s,
\end{eqnarray}
which transform under gauge transformation as
\begin{eqnarray}
\delta_V C_{(3)} &=& \mathcal{L}_vC_{(3)}+d\tilde{v}_{(2)},\\
\delta_V C_{(6)} &=& \mathcal{L}_vC_{(6)}+d\tilde{v}_{(5)}-\frac{1}{2}C_{(3)}\wedge d\tilde{v}_{(2)}.
\end{eqnarray}
This allows us to define $\hat{w}_{(2)}$ and $\hat{w}_{(5)}$ by
\begin{eqnarray}
\hat{w}_{(2)} &=& \tilde{w}_{(2)}+\iota_wC_{(3)},\\
\hat{w}_{(5)} &=& \tilde{w}_{(5)}+\tilde{w}_{(2)}\wedge C_{(3)}+\frac{1}{2}\iota_w C_{(3)}\wedge C_{(3)}+\iota_wC_{(6)}.
\end{eqnarray}
Under a gauge transformation, these objects transform as a 2-form and a 5-form, respectively,
\begin{eqnarray}
\delta_V\hat{w}_{(2)} &=& \mathcal{L}_v \hat{w}_{(2)},\\
\delta_V\hat{w}_{(5)} &=& \mathcal{L}_v \hat{w}_{(5)},
\end{eqnarray} 
and are invariant under the $\tilde{v}$ transformation.
Therefore, $\hat{W} = {w}\oplus\hat{w}_{(2)}\oplus\hat{w}_{(5)}$ is a section of $(T\oplus \Lambda^2 T^*\oplus \Lambda^5 T^*) U$. Their finite transformations are given by
\begin{eqnarray}
{w}'(x') &=& {w}(x),\\
\hat{w}_{(2)}'(x') &=& \hat{w}_{(2)}(x),\\
\hat{w}_{(5)}'(x') &=& \hat{w}_{(5)}(x),
\end{eqnarray}
where $x' = e^{-v^m\partial_m}x$.

Using the finite transformation of the coordinate bases given by
\be
e'_m(x') = e_n(x)\frac{\partial x^n}{\partial x'^m}  \qquad e'^m(x') = e^n(x)\frac{\partial x'^m}{\partial x^n},
\ee
the finite transformations of the components of the $\hat{W}$  can be written as
\begin{eqnarray}
{w}'^m(x') &=& w^n(x)\frac{\partial x'^m}{\partial x^n},  \\
\hat{w}'_{mn}(x') &=& \hat{w}_{pq}(x)\frac{\partial x^p}{\partial x'^m}\frac{\partial x^q}{\partial x'^n},\\
\hat{w}'_{mnpqr}(x') &=& \hat{w}_{stuvw}(x)\frac{\partial x^s}{\partial x'^m} \frac{\partial x^t}{\partial x'^n} \frac{\partial x^u}{\partial x'^p} \frac{\partial x^v}{\partial x'^q} \frac{\partial x^w}{\partial x'^r}.
\end{eqnarray}
The finite transformation of the gerbe connections can be taken to be
\begin{eqnarray}
C'_{(3)}(x') &=& C_{(3)}(x)+d\tilde{v}_{(2)}(x),\\
C_{(6)}'(x') &=& C_{(6)}(x)+d\tilde{v}_{(5)}(x)-\frac{1}{2}C_{(3)}(x)\wedge d\tilde{v}_{(2)}(x),
\end{eqnarray}
so that
\begin{eqnarray}
\hat{w}_{(2)}'(x') &=& \tilde{w}_{(2)}'(x')+\iota_{w'}C'_{(3)}(x')\\
\hat{w}_{(5)}'(x') &=& \tilde{w}_{(5)}'(x')+\tilde{w}_{(2)}'(x')\wedge C'_{(3)}(x')+\frac{1}{2}\iota_{w'}C'_{(3)}(x')\wedge C'_{(3)}(x')+\iota_{w'}C_{(6)}'(x').
\end{eqnarray}
This then gives the finite transformation of $\tilde{w}_{(2)}$:
\begin{eqnarray}
\tilde{w}_{(2)}'(x') &=& \hat{w}'_{(2)}(x')-\iota_{w'}C'_{(3)}(x').\nonumber\\
&=& \hat{w}_{(2)}(x)-\iota_w(C_{(3)}(x)+d\tilde{v}_{(2)}(x)),\nonumber\\
&=& \hat{w}_{(2)}(x)-\iota_wC_{(3)}(x)-\iota_wd\tilde{v}_{(2)}(x),\nonumber\\
&=& \tilde{w}_{(2)}(x)-\iota_wd\tilde{v}_{(2)}(x),
\end{eqnarray}
and the finite transformation of $\tilde{w}_{(5)}$:
\begin{eqnarray}
\tilde{w}_{(5)}'(x') &=& \hat{w}_{(5)}'(x')-\tilde{w}_{(2)}'(x')\wedge C'_{(3)}(x')-\frac{1}{2}\iota_{w'}C'_{(3)}(x')\wedge C'_{(3)}(x')-\iota_{w'}C_{(6)}'(x'),\nonumber\\
&=& (\tilde{w}_{(5)}(x)+\tilde{w}_{(2)}(x)\wedge C_{(3)}(x)+\frac{1}{2}\iota_w C_{(3)}(x)\wedge C_{(3)}(x)+\iota_w\tilde{C_{(3)}}(x))\nonumber\\
& &-(\tilde{w}_{(2)}(x)-\iota_wd\tilde{v}_{(2)}(x))\wedge(C_{(3)}(x)+d\tilde{v}_{(2)}(x))\nonumber\\
& &-\frac{1}{2}\iota_w(C_{(3)}(x)+d\tilde{v}_{(2)}(x))\wedge(C_{(3)}(x)+d\tilde{v}_{(2)}(x))\nonumber\\
& &-i_{w}(C_{(6)}(x)+d\tilde{v}_{(5)}(x)-\frac{1}{2}C_{(3)}(x)\wedge d\tilde{v}_{(2)}(x)),\nonumber\\
&=& \tilde{w}_{(5)}(x)-\tilde{w}_{(2)}(x)\wedge d\tilde{v}_{(2)}(x)+\frac{1}{2}\iota_wd\tilde{v}_{(2)}(x)\wedge d\tilde{v}_{(2)}(x)-\iota_wd\tilde{v}_{(5)}(x).
\end{eqnarray}

In summary, the transformation of $W$ is given by
\begin{eqnarray}
w'(x') &=& w(x),\\
\tilde{w}_{(2)}'(x') &=& \tilde{w}_{(2)}(x)-\iota_wd\tilde{v}_{(2)}(x),\\
\tilde{w}_{(5)}'(x') &=& \tilde{w}_{(5)}(x)-\tilde{w}_{(2)}(x)\wedge d\tilde{v}_{(2)}(x)+\frac{1}{2}\iota_wd\tilde{v}_{(2)}(x)\wedge d\tilde{v}_{(2)}(x)-\iota_wd\tilde{v}_{(5)}(x),
\end{eqnarray}
which implies the components $(w^m, \tilde{w}_{mn}, \tilde{w}_{mnpqr})$ transform as
\begin{eqnarray}
w'^m(x') &=& w^n(x)\frac{\partial x'^m}{\partial x^n},\\
\tilde{w}'_{mn}(x') &=& \left(\tilde{w}_{pq}(x)-3w^r(x)(d\tilde{v}_{(2)})_{rpq}(x)\right)\frac{\partial x^p}{\partial x'^m} \frac{\partial x^q}{\partial x'^n},\\
\tilde{w}'_{mnpqr}(x') &=&\left(\tilde{w}_{stuvw}(x)-30\tilde{w}_{[st}(x)(d\tilde{v}_{(2)})_{uvw]}(x)+15w^l(x)(d\tilde{v}_{(2)})_{l[st}(x)(d\tilde{v}_{(2)})_{uvw]}(x)\right.\nonumber\\
& & \left.-6w^l(x)(d\tilde{v}_{(5)})_{lstuvw}(x)\right)\frac{\partial x^s}{\partial x'^m} \frac{\partial x^t}{\partial x'^n} \frac{\partial x^u}{\partial x'^p} \frac{\partial x^v}{\partial x'^q} \frac{\partial x^w}{\partial x'^r},
\end{eqnarray}
where $(d\tilde{v}_{(2)})_{rpq} = \partial_{[r}\tilde{v}_{pq]}$ and $(d\tilde{v}_{(5)})_{mnpqrs} = \partial_{[m}\tilde{v}_{npqrs]}$.


\section{Generalised Tensors}  \setcounter{equation}{0}

In this section, we review the construction of tensors and twisted tensors in DFT of \cite{Hull:2014mxa} and then generalise this to EFT .

\subsection{ Generalised Tensors of Double Field Theory}

A generalised vector $W^M$ tranforms as a vector under $O(D,D)$,
so that under $GL(D,\mathbb {R})\subset O(D,D)$
it transforms reducibly as
\be
W\to \hat R(\lambda) W
\ee
where  $\lambda ^m{}_n \in GL(D,\mathbb {R}) $ and
\be
\label{hatris}
\hat R (\lambda)= 
\begin{pmatrix}
\lambda  &0 \\
0& (\lambda ^{-1})^t  \end{pmatrix}\,
\ee
The untwisted version  of $W$ is $\hat W^M$,  which  can be written as
\be
\hat W = L W
\ee
where
\be
\label{liss}
L= 
\begin{pmatrix}
1 &0 \\
-B &1  \end{pmatrix}\,
\ee
 denotes the  matrix with components
\be
L^M{}_N = \begin{pmatrix}
\d^m{}_n &0 \\
-B_{mn}
&\d _m {}^n  \end{pmatrix}\,.
\ee
The transformation of the  untwisted vector $\hat W$  is then
\be
\hat W' (X') =\hat R(\L) \hat W(X)
\ee
where  
\be
\L ^m {}_n =\frac{
  \partial x'{}^m }
  {\partial x^n}.
  \ee
The coordinate transformation acts only on the $x$:
  \be
X^M \to X'{}^M= \begin{pmatrix} x'{}^m \\   \tilde x'_m \end{pmatrix} \, ,
\ee
with
\be
x^m \to x'{}^m(x), \qquad \ti x_m \to \ti x'_m= \ti x_m
\ee

  The transformation of the twisted vector $W$ was found by twisting the untwisted transformation and is
 \be
 \label{Rtranss}
  W' (X') =  R   W(X)
\ee 
where
\be 
\label{risss}
R= L'(X')^{-1}\hat R(\Lambda) L(X) = \hat R (\Lambda) S
\ee
and
\be
L'(X')= 
\begin{pmatrix}
1 &0 \\
-B'(x') &1  \end{pmatrix}\,
\ee
with $B'(x')$ given by (\ref{BTranss}), and 
\be
\label{radss}
S=  \begin{pmatrix}
\d^m{}_n&0 \\
2\partial _{[m}\ti v_{n]} 
&\d _m {}^n \end{pmatrix}\,.
\ee
The matrices $R,\hat R, L , S$ are all in $O(D,D)$.

Lowering indices with $\eta$ gives similar formulae for a generalised vector with lower index
 \be
U_M = \begin{pmatrix}   \tilde  u_m \\ u^m \end{pmatrix} \, .
\ee
The untwisted vector
\be
\hat U_M =
 \begin{pmatrix}   \hat u_m \\ u^m \end{pmatrix}
 =
  \begin{pmatrix}   \tilde u_m- B_{mn} u^n
 \\ u^m \end{pmatrix}
\ee
transforms with 
\be
 \delta \hat U_M=  {\cal L}_{v}
 \hat U_M
 \ee
and is invariant under $\tilde v$ transformations.
Then the untwisted vector is
\be
\hat U= 
 U L^{-1}
\ee
(i.e. $\hat U_M = U_N (L^{-1})^N {}_M$; recall
$ \eta L \eta ^{-1} = (L^t)^{-1}
$ as $L$ is in $O(D,D)$)
 and transforms under a finite transformation as
 \be
 \hat U'(X') = \hat U(X) \hat R^{-1}
 \ee
 where here and in what follows
$\hat R=\hat R (\Lambda)$.
For the twisted vectors
 \be
   U'(X') =      U(X) R^{-1}
 \ee

This extends to arbitrary generalised tensors
$T^{MN\dots }{}_{PQ\dots}$.
We define the untwisted tensor
\be 
\hat T^{MN\dots }{}_{PQ\dots} = L^M{}_R L^N{}_S   \dots 
T^{RS\dots }{}_{TU\dots}
  ( L ^{-1})
^T {} _P{}
(   L ^{-1})
^U{}  _Q \dots \label{untwisted}
\ee
which transforms as
\be 
\hat T' {} ^{MN\dots }{}_{PQ\dots} (X')= \hat R ^M{}_R  \hat R ^N{}_S   \dots T^{RS\dots }{}_{TU\dots}  (
 \hat R ^{-1})
^T {} _P{}
( \hat R ^{-1})
^U{}  _Q \dots 
\ee
so that the original tensor transforms as
\be 
  T' {} ^{MN\dots }{}_{PQ\dots} (X')=   R ^M{}_R    R ^N{}_S   \dots T^{RS\dots }{}_{TU\dots} 
   ( R ^{-1})
^T {} _P{}
(   R ^{-1})
^U{}  _Q \dots 
\ee
Raising all  lower indices with $\eta$ gives a generalised tensor
$T^{M_1\dots M_p}$ of some rank $p$ which is a section of  $E^p$ while
$\hat T^{M_1\dots M_p}$ is a section of $(T\oplus T^*)^p$.
In particular, 
\be \hat \eta _{MN}= \eta_{MN}
\ee
as $L\in O(D,D)$, and is invariant, $\eta ' = \eta$.


\subsection{ Generalised Tensors of Extended Field Theory}

For each of the extended field theories, a similar structure applies, with matrices $L,L',R,\hat R,S$, all of which are in the duality group $G$ which is $SL(5,\mathbb {R})$, $SO(5,5)$
or $E_6$.
The untwisted form $\hat W^M$ of a generalised vector $W^M$
can be written as
\be
\hat W = L W.
\ee
The generalised vector  transforms as a representation of  $G$ and decompose into a reducible representation of $GL(d,\mathbb {R})$ (where $d$ is the rank of $G$) under which $\lambda \in GL(d,\mathbb {R}) $ acts on $W$ to give $W\to \hat R(\lambda) W$. 
The transformation of the  untwisted vector $\hat W$  is then
\be
\hat W' (X') =\hat R(\lambda)  \hat W(X)
\ee
where  
\be
\L ^m {}_n =\frac{
  \partial x'{}^m }
  {\partial x^n}.
  \ee
The transformation of the twisted vector $W$ was found by twisting the untwisted transformation and can be written as
 \be
 \label{Rtranss}
  W' (X') =  R   W(X)
\ee 
where
\be 
\label{risss}
R= L'(X')^{-1}\hat R (\Lambda) L(X) = \hat R(\Lambda) S
\ee
where $L'$ generates shifts of the antisymmetric tensor gauge fields and $S$ is the corresponding gauge transformation; see below for explicit forms.

As for the DFT case, this extends to arbitrary generalised tensors
$T^{MN\dots }{}_{PQ\dots}$.
We define the untwisted tensor
\be 
\hat T^{MN\dots }{}_{PQ\dots} = L^M{}_R L^N{}_S   \dots 
T^{RS\dots }{}_{TU\dots}
  ( L ^{-1})
^T {} _P{}
(   L ^{-1})
^U{}  _Q \dots 
\ee
which transforms as
\be 
\hat T' {} ^{MN\dots }{}_{PQ\dots} (X')= \hat R ^M{}_R  \hat R ^N{}_S   \dots T^{RS\dots }{}_{TU\dots}  (
 \hat R ^{-1})
^T {} _P{}
( \hat R ^{-1})
^U{}  _Q \dots 
\ee
so that the original tensor transforms as
\be 
  T' {} ^{MN\dots }{}_{PQ\dots} (X')=   R ^M{}_R    R ^N{}_S   \dots T^{RS\dots }{}_{TU\dots} 
   ( R ^{-1})
^T {} _P{}
(   R ^{-1})
^U{}  _Q \dots 
\ee

We now give the explicit forms of the matrices appearing above.
For the  $SL(5,\mathbb {R})$ case, the untwisted vector is
\be
\hat{W}^M  = L^M{}_N W^N
= \begin{pmatrix}
\d^m{}_l &0 \\
C_{lmn}
&\d _{mn} {}^{pq}  \end{pmatrix} \begin{pmatrix}   w^l \\ \tilde{w}_{pq}\end{pmatrix} \, ,
\ee
where $\delta_{mn}{}^{pq} = \frac{1}{2}(\delta_m{}^p \delta_n{}^q-\delta_m{}^q\delta_n{}^p)$.
For the $SO(5,5)$ case, the untwisted vector is
\be
\hat W^M = L^M{}_N W^N = \begin{pmatrix}
\d^m{}_l &0 &0 \\
C_{lmn} &\d _{mn} {}^{pq}& 0\\
5C_{l[mn}C_{pqr]} & 10\d_{[mn}{}^{pq}C_{rst]} & \d_{mnpqr}{}^{stuvw}
  \end{pmatrix} \begin{pmatrix}   w^l \\ \tilde{w}_{pq} \\ \tilde{w}_{stuvw}\end{pmatrix},
\ee
where $\d_{mnpqr}{}^{stuvw} = \d_{[m}{}^s\dots\d_{r]}{}^{w}$.
For $E_6$ case, the untwisted vector is
\be
\hat W^M = L^M{}_N W^N = \begin{pmatrix}
\d^m{}_l &0 &0 \\
C_{lmn} &\d _{mn} {}^{pq}& 0\\
{C}_{lmnpqr}+5C_{l[mn}C_{pqr]} & 10\d_{[mn}{}^{pq}C_{rst]} & \d_{mnpqr}{}^{stuvw}
  \end{pmatrix} \begin{pmatrix}   w^l \\ \tilde{w}_{pq} \\ \tilde{w}_{stuvw}\end{pmatrix} 
\ee
Then the  $L $ matrix for the $SL(5,\mathbb {R})$ theory is
\be \label{L4}
L ^M{}_N   = \begin{pmatrix}
\d^m{}_l &0 \\
C _{lmn}  
&\d _{mn} {}^{pq}  \end{pmatrix}\, ,
\ee
while for the $SO(5,5)$ theory it is
\be  \label{L5}
L ^M{}_N   = \begin{pmatrix}
\d^m{}_l &0 &0 \\
C _{lmn}   &\d _{mn} {}^{pq}& 0\\
5C _{l[mn}  C _{pqr]}   & 10\d_{[mn}{}^{pq}C _{rst]}   & \d_{mnpqr}{}^{stuvw}
  \end{pmatrix}\,
\ee
and for the $E_6$ theory it is
\be  \label{L6}
L ^M{}_N   = \begin{pmatrix}
\d^m{}_l &0 &0 \\
C _{lmn}   &\d _{mn} {}^{pq}& 0\\
{C} _{lmnpqr}  +5C _{l[mn}  C _{pqr]}   & 10\d_{[mn}{}^{pq}C _{rst]}   & \d_{mnpqr}{}^{stuvw}
  \end{pmatrix}\, .
\ee

 The matrices   $\hat R$ for the
 $SL(5,\mathbb {R}),SO(5,5)$
and $E_6$  theories are given by
 \be
\hat R^M{}_N= \begin{pmatrix}
\Lambda^m{}_l &0 \\
0
&(\Lambda^{-1})_m{}^p(\Lambda^{-1})_n{}^q  \end{pmatrix},
\ee

 \be
\hat R^M{}_N= \begin{pmatrix}
\Lambda^m{}_l &0 &0 \\
0 &(\Lambda^{-1})_m{}^p(\Lambda^{-1})_n{}^q & 0\\
 0 & 0 & (\Lambda^{-1})_m{}^s (\Lambda^{-1})_n{}^t (\Lambda^{-1})_p{}^u (\Lambda^{-1})_q{}^v (\Lambda^{-1})_r{}^w  \end{pmatrix},
\ee

 \be
\hat R^M{}_N= \begin{pmatrix}
\Lambda^m{}_l &0 &0 \\
0 &(\Lambda^{-1})_m{}^p(\Lambda^{-1})_n{}^q & 0\\
 0 & 0 & (\Lambda^{-1})_m{}^s (\Lambda^{-1})_n{}^t (\Lambda^{-1})_p{}^u (\Lambda^{-1})_q{}^v (\Lambda^{-1})_r{}^w  \end{pmatrix},
\ee
 respectively.

The $L'$ matrix for the $SL(5,\mathbb {R})$ theory is
\be
L'^M{}_N(X') = \begin{pmatrix}
\d^m{}_l &0 \\
C'_{lmn}(X')
&\d _{mn} {}^{pq}  \end{pmatrix}\, ,
\ee
while for the $SO(5,5)$ theory it is
\be
L'^M{}_N(X') = \begin{pmatrix}
\d^m{}_l &0 &0 \\
C'_{lmn}(X') &\d _{mn} {}^{pq}& 0\\
5C'_{l[mn}(X')C'_{pqr]}(X') & 10\d_{[mn}{}^{pq}C'_{rst]}(X') & \d_{mnpqr}{}^{stuvw}
  \end{pmatrix}\,
\ee
and for the $E_6$ theory it is
\be
L'^M{}_N(X') = \begin{pmatrix}
\d^m{}_l &0 &0 \\
C'_{lmn}(X') &\d _{mn} {}^{pq}& 0\\
{C}'_{lmnpqr}(X')+5C'_{l[mn}(X')C'_{pqr]}(X') & 10\d_{[mn}{}^{pq}C'_{rst]}(X') & \d_{mnpqr}{}^{stuvw}
  \end{pmatrix}\, .
\ee

Finally, the matrix $S$ for the $SL(5,\mathbb {R})$ theory is
\be
S^M{}_N = \begin{pmatrix}
\d^m{}_l &0 \\
-3(d\tilde{v}_{(2)})_{lmn} &\d _{mn} {}^{pq}  \end{pmatrix}\,
\ee
while for the $SO(5,5)$ theory it is
\be
S^M{}_N = \begin{pmatrix}
\d^m{}_l &0 &0 \\
-3(d\tilde{v}_{(2)})_{lmn} &\d _{mn} {}^{pq}& 0\\
15(d\tilde{v}_{(2)})_{l[mn} (d\tilde{v}_{(2)})_{pqr]}  & -30\d_{[mn}{}^{pq}\partial_r\tilde{v}_{st]} & \d_{mnpqr}{}^{stuvw}
\end{pmatrix}\,
\ee
and for the $E_6$ theory it is
\be
S^M{}_N = \begin{pmatrix}
\d^m{}_l &0 &0 \\
-3(d\tilde{v}_{(2)})_{lmn} &\d _{mn} {}^{pq}& 0\\
-6(d\tilde{v}_{(5)})_{lmnpqr} +15(d\tilde{v}_{(2)})_{l[mn} (d\tilde{v}_{(2)})_{pqr]} & -30\d_{[mn}{}^{pq}\partial_r\tilde{v}_{st]} & \d_{mnpqr}{}^{stuvw}
  \end{pmatrix}\,,
\ee
where $(d\tilde{v}_{(2)})_{rpq} = \partial_{[r}\tilde{v}_{pq]}$ and $(d\tilde{v}_{(5)})_{mnpqrs} = \partial_{[m}\tilde{v}_{npqrs]}$.

\section{The Generalised Metric in DFT and  EFT} \setcounter{equation}{0}

For DFT and EFT, there is a duality group $G$
(which is $O(D,D)$ for DFT and $E_D$ for EFT) with maximal compact subgroup $H$.
Remarkably, the fields $(g_{mn}, B_{mn})$ in DFT and 
$(g_{mn}, C_{mnp},C_{mnpqrs})$ in EFT
can be regarded as a field taking values in the coset $G/H$  -- the coset space can be locally parameterised by $g_{mn}, B_{mn}$ or $g_{mn}, C_{mnp},C_{mnpqrs}$ \cite{Hull:2007zu}.
A field taking values in the coset $G/H$ can be represented by a vielbein ${\cal V}^A{}_M(X)\in G$ transforming
as
\be
{\cal V} \to h{\cal V}g
\ee
under a rigid $G$ transformation $ g^N{}_M\in G$ and a local $H$ transformation
$h^A{}_B(X)\in H$.
If $k_{AB}$ is an $H$-invariant metric ($h^tkh=k $ for all $h\in H$)
then the degrees of freedom can also be encoded in a generalised metric
\be
\mathcal{H}_{MN} = k_{AB}\, {\cal V}^A{}_M {\cal V}^B{}_N
\ee
which by construction is invariant under  $H$ transformations.

We now show how the coset is parameterised by the fields $(g_{mn}, B_{mn})$ or
$(g_{mn}, C_{mnp},C_{mnpqrs})$.
Let $e^a{}_m$ be a vielbein for the metric $g_{mn}$, with
\be
g_{mn} = \delta _{ab}\, e^a{}_me^b{}_n
\ee
and inverse vielbein $e_a{}^m$. The indices $a,b$ transform under the tangent space group $O(D)$.
Then the vielbein ${\cal V}$ can be written in terms of $e^a{}_m$ and $(g_{mn}, B_{mn})$ or
$(g_{mn}, C_{mnp},C_{mnpqrs})$
as
\be
{\cal V} = h\hat R(e^a{}_m)L
\ee
(This can be viewed as  a consequence of the Iwasawa decomposition theorem.)
Here $h\in H$ and can be chosen to be $h=1$ by a local $H$ transformation.
The dependence on the gauge fields is given by the matrix $L(B)$ in DFT and $L(
C_{(3)},
C_{(6)})$
in EFT; $L$ is of the form $L=1+N$ where $N$ is lower triangular.
Finally $\hat R(e) $ is the matrix $\hat R(\lambda) $ given above, with $\lambda ^a{}_m = e^a{}_m$ and serves to convert \lq curved' indices $m,n$ to  \lq flat' indices $a,b$.

Then the generalised metric is given by
\be
\mathcal{H} (V,W)= \hat {\mathcal{H}}  ( \hat V,\hat W)
\ee
 for any generlised vectors $V,W$, 
 where
 \be
 \hat 
 {\mathcal{H}}
 (\hat V,\hat W)
= k (\hat R(e) \hat V,\hat R(e)\hat W)
\ee
Explicitly,
\be  \hat 
 {\mathcal{H}}_{MN} = k_{AB} \hat R(e) ^A{}_M \hat R(e) ^B{}_N
\ee
and
\be \label{histtt}
\mathcal{H}_{MN}
= \hat{ \mathcal{H}}_{PQ}L^P{}_M L^Q{}_N
\ee
Then $\hat {\mathcal{H}}_{PQ}$ is the untwisted form of the generalised metric, and is the natural metric on generalised or extended tangent vectors induced by the metric $g$ for ordinary vectors.

We now show how this works for the cases discussed here.
For DFT, we recover the discussion of \cite{Hull:2014mxa}. An untwisted vector decomposes as
\be
\hat W^M = \begin{pmatrix}   w^m \\ \hat w_m \end{pmatrix} \, ,
\ee
so
\be
\hat R(e)\hat W  = \begin{pmatrix}   w^a \\ \hat w_a \end{pmatrix} 
\ee
where as usual
\be
w^a= e^a{}_m w^m, \qquad
\hat w_a = e_a{}^m
\hat w_m
\ee
The metric $k_{AB}$ decomposes under $O(D)$ to give
\be\label{Hisst}
k_{AB} \ = \  \begin{pmatrix}    
  \delta_{ab}   & 
0 \\  0   &
  \delta ^{ab}
   \end{pmatrix}\;
 \ee
 so
 \be
 k (\hat R(e) \hat V,\hat R(e)\hat W)
 = \delta_{ab} v^a w^b + \delta ^{ab} \hat v_a \hat w_b
\ee
This is equal to $\hat 
 {\mathcal{H}}
 (\hat V,\hat W)=\hat{\mathcal{H}}_{MN}\hat V^M \hat W^N$ so
\be\label{Hisst}
 \hat  {\cal H}_{MN} \ = \  \begin{pmatrix}    
  g_{mn}   & 
0 \\  0   &
  g^{mn}
   \end{pmatrix}\;
 \ee
 which is the standard metric on $T\oplus T^*$ induced by the metric $g$ on $T$.
 Then the generalised metric is
  \be\label{Hisskoit}
  {\cal H}_{MN} \ = \   \hat {\cal H}_{PQ} 
   L^P {} _M{}
L
^Q{}  _N 
  \ee
where $L(B)$ is given by (\ref{liss}).
This gives the standard result
\be\label{His}
  {\cal H}_{MN} \ = \  \begin{pmatrix}    
  g_{mn}-B_{mk}g^{kl}B_{ln}  & 
 B_{mk}g^{kn}   \\[0.5ex]
 -g^{mk}B_{kn}   &
  g^{mn}
   \end{pmatrix}   
   \;
 \ee

Consider now EFT with $G= E_D$ for $D=5,6$.
An untwisted vector decomposes as
\be
\hat{W}^M = \begin{pmatrix}   w^m \\ \hat {w}_{mn} \\ \hat{w}_{mnpqr} \end{pmatrix} \, ,
\ee
so
\be
\hat R(e)\hat W  = \begin{pmatrix}   w^a \\ \hat {w}_{ab} \\ \hat{w}_{abcde} \end{pmatrix} \, ,
\ee
with all indices converted to tangent space indices using $e^a{}_m$.
The metric $k_{AB}$ decomposes under $O(D)$ to give \be
 k (\hat R(e) \hat V,\hat R(e)\hat W)
 =
 v^a w_a + \frac 1 2  
  \hat {v}_{ab}\hat {w}^{ab} + \frac 1 {5!}  \hat{v}_{abcde} \hat{w}^{abcde}
\ee
where indices have been raised or lowered with $\delta_{ab} $.
The matrix can then be written as
\be\label{Hisstp}
k_{AB} \ = \  \begin{pmatrix}    
  \delta_{ab}   & 
0 \\  0   &
  \delta ^{ab,cd}
   & 
0 \\  
0 & 
0 & \delta ^{a_1 \dots a_5,b_1 \dots b_5}
\end{pmatrix}\; 
 \ee
 where 
 $\delta ^{ab,cd} =\frac 1 2  \delta ^{c[a }\delta ^{b]d}$ and a similar expression for $\delta ^{a_1 \dots a_5,b_1 \dots b_5}$.
Then
 (\ref{Hisstp})
is equal to $\hat 
 {\mathcal{H}}
 (\hat V,\hat W)=\hat{\mathcal{H}}_{MN}\hat V^M \hat W^N$ so
\be
\hat {\mathcal{H}}_{MN}\hat V^M \hat W^N
=v^ m w_ m + \frac 1 2  
  \hat {v}_{ m n}\hat {w}^{ m n} + \frac 1 {5!}  \hat{v}_{ m npqr} \hat{w}^{ m npqr}
\ee
where indices have been raised and lowered with $g_{mn}$.
The corresponding matrix is
\be\label{Hisstsdf}
 \hat  {\cal H}_{MN} \ = \  \begin{pmatrix}    
  g_{mn}  & 
0 \\  0   &
  g ^{mn,pq}
   & 
0 \\  
0 & 
0 & g ^{m_1 \dots m_5,n_1 \dots n_5}
\end{pmatrix}\; 
 \ee
 where  
 $g ^{lk,mn} =\frac 1 2  g ^{m[l }g ^{k]n}$ and a similar expression for $g ^{a_1 \dots a_5,b_1 \dots b_5}$.
 Then $ \hat  {\cal H}_{MN}$ is the standard metric on $T\oplus \Lambda^2 T^*\oplus \Lambda^5 T^*$  induced by the metric $g$ on $T$.

The generalised metric is then the twisted from of this
  \be\label{Hisskoit}
  {\cal H}_{MN} \ = \   \hat {\cal H}_{PQ} 
   L^P {} _M{}
L
^Q{}  _N 
  \ee
where  $L(
C_{(3)},
C_{(6)})$
is given by    (\ref{L5}) or  (\ref{L6}). This then gives explicit forms for the generalised metric, in agreement with   \cite{Hull:2007zu,Berman:2011pe,Hohm:2013vpa,Musaev:2015pla,Abzalov:2015ega,Berman:2011jh,Baguet:2015xha}.

For $SL(5,\mathbb{R})$, there is no $C_{(6)}$ or $\hat{w}_{(5)}$, but similar fomulae apply with
\be
\hat{\mathcal{H}}_{MN} = \begin{pmatrix}
g_{mn}& 0\\
0 &
g^{kl,pq}  \end{pmatrix}\, .
\ee
and (\ref{histtt}) with $L(
C_{(3)}$ given by 
(\ref{L4}) gives
the 
\be
\mathcal{H}_{MN} = \begin{pmatrix}
g_{mn}+
C_m{}^{rs}C_{rsn} &  \frac{1}{2}C_m{}^{pq} \\
\frac{1}{2}C^{kl}{}_{n}   &
g^{kl,pq}  \end{pmatrix}\, ,
\ee
recovering   the generalised metric given in \cite{Hull:2007zu}.

\section*{Acknowledgement}

The work of CMH was supported  by STFC grant ST/L00044X/1 and EPSRC grant EP/K034456/1.


\begin{thebibliography}{99}

\baselineskip 15pt






\bibitem{Hull:2009mi}
  C.~Hull, B.~Zwiebach,
  ``Double Field Theory,''
  JHEP {\bf 0909}, 099 (2009).
  [arXiv:0904.4664 [hep-th]],


\bibitem{Hohm:2010jy}
  O.~Hohm, C.~Hull and B.~Zwiebach,
  ``Background independent action for double field theory,''
  JHEP {\bf 1007} (2010) 016
  [arXiv:1003.5027 [hep-th]].

\bibitem{Hohm:2010pp}
  O.~Hohm, C.~Hull and B.~Zwiebach,
  ``Generalised metric formulation of double field theory,''
  JHEP {\bf 1008} (2010) 008
  [arXiv:1006.4823 [hep-th]].

\bibitem{Siegel:1993th}
  W.~Siegel,
  ``Superspace duality in low-energy superstrings,''
  Phys.\ Rev.\  D {\bf 48}, 2826 (1993)
  [arXiv:hep-th/9305073]. 

\bibitem{Hitchin:2004ut} 
  N.~Hitchin,
  ``Generalised Calabi-Yau manifolds,''
  Quart.\ J.\ Math.\ Oxford Ser.\  {\bf 54}, 281 (2003)
  [math/0209099 [math-dg]].
  
\bibitem{Gualtieri:2003dx} 
  M.~Gualtieri,
  ``Generalised complex geometry,''
  math/0401221 [math-dg].
  
\bibitem{Hitchin:2005in}
  N.~Hitchin,
  ``Brackets, forms and invariant functionals,''
  math/0508618 [math-dg].
  
\bibitem{Hitchin:2010qz}
  N.~Hitchin,
  ``Lectures on generalized geometry,''
  arXiv:1008.0973 [math.DG].

\bibitem{Grana:2008yw}
  M.~Grana, R.~Minasian, M.~Petrini and D.~Waldram,
  ``T-duality, Generalized Geometry and Non-Geometric Backgrounds,''
  JHEP {\bf 0904} (2009) 075
  [arXiv:0807.4527 [hep-th]].
  
\bibitem{Coimbra:2011nw}
  A.~Coimbra, C.~Strickland-Constable and D.~Waldram,
  ``Supergravity as Generalised Geometry I: Type II Theories,''
  JHEP {\bf 1111} (2011) 091
  [arXiv:1107.1733 [hep-th]].
  
\bibitem{Lee:2015xga}
  K.~Lee, C.~Strickland-Constable and D.~Waldram,
  ``New gaugings and non-geometry,''
  arXiv:1506.03457 [hep-th].

\bibitem{Hull:2014mxa}
  C.~M.~Hull,
  ``Finite Gauge Transformations and Geometry in Double Field Theory,''
  JHEP {\bf 1504} (2015) 109
  doi:10.1007/JHEP04(2015)109
  [arXiv:1406.7794 [hep-th]].

\bibitem{Aldazabal:2013sca}
  G.~Aldazabal, D.~Marques and C.~Nunez,
  ``Double Field Theory: A Pedagogical Review,''
  Class.\ Quant.\ Grav.\  {\bf 30} (2013) 163001
  [arXiv:1305.1907 [hep-th]].
  
\bibitem{Berman:2013eva}
  D.~S.~Berman and D.~C.~Thompson,
  ``Duality Symmetric String and M-Theory,''
  arXiv:1306.2643 [hep-th].
  
\bibitem{Hohm:2013bwa}
  O.~Hohm, D.~LŸst and B.~Zwiebach,
  ``The Spacetime of Double Field Theory: Review, Remarks, and Outlook,''
  Fortsch.\ Phys.\  {\bf 61} (2013) 926
  [arXiv:1309.2977 [hep-th]].

  

\bibitem{Hull:2007zu}
  C.~M.~Hull,
  ``Generalised Geometry for M-Theory,''
  JHEP {\bf 0707} (2007) 079
  doi:10.1088/1126-6708/2007/07/079
  [hep-th/0701203].

\bibitem{Pacheco:2008ps}
  P.~P.~Pacheco and D.~Waldram,
  ``M-theory, exceptional generalised geometry and superpotentials,''
  JHEP {\bf 0809} (2008) 123
  doi:10.1088/1126-6708/2008/09/123
  [arXiv:0804.1362 [hep-th]].
  
\bibitem{Hull:2004in}
  C.~M.~Hull,
  ``A Geometry for non-geometric string backgrounds,''
  JHEP {\bf 0510} (2005) 065
  [hep-th/0406102].

\bibitem{Berman:2010is}
  D.~S.~Berman and M.~J.~Perry,
  ``Generalized Geometry and M theory,''
  JHEP {\bf 1106} (2011) 074
  doi:10.1007/JHEP06(2011)074
  [arXiv:1008.1763 [hep-th]].
  
\bibitem{Berman:2011pe}
  D.~S.~Berman, H.~Godazgar and M.~J.~Perry,
  ``SO(5,5) duality in M-theory and generalized geometry,''
  Phys.\ Lett.\ B {\bf 700} (2011) 65
  doi:10.1016/j.physletb.2011.04.046
  [arXiv:1103.5733 [hep-th]].

\bibitem{Berman:2011cg}
  D.~S.~Berman, H.~Godazgar, M.~Godazgar and M.~J.~Perry,
  ``The Local symmetries of M-theory and their formulation in generalised geometry,''
  JHEP {\bf 1201} (2012) 012
  doi:10.1007/JHEP01(2012)012
  [arXiv:1110.3930 [hep-th]].
  
\bibitem{Cederwall:2013naa}
  M.~Cederwall, J.~Edlund and A.~Karlsson,
  ``Exceptional geometry and tensor fields,''
  JHEP {\bf 1307} (2013) 028
  doi:10.1007/JHEP07(2013)028
  [arXiv:1302.6736 [hep-th]].
  
\bibitem{Hohm:2013pua}
  O.~Hohm and H.~Samtleben,
  ``Exceptional Form of D=11 Supergravity,''
  Phys.\ Rev.\ Lett.\  {\bf 111} (2013) 231601
  doi:10.1103/PhysRevLett.111.231601
  [arXiv:1308.1673 [hep-th]].

\bibitem{Hohm:2013vpa}
  O.~Hohm and H.~Samtleben,
  ``Exceptional Field Theory I: $E_{6(6)}$ covariant Form of M-Theory and Type IIB,''
  Phys.\ Rev.\ D {\bf 89} (2014) 6,  066016
  doi:10.1103/PhysRevD.89.066016
  [arXiv:1312.0614 [hep-th]].
 
  
\bibitem{Hohm:2013uia}
  O.~Hohm and H.~Samtleben,
  ``Exceptional field theory. II. E$_{7(7)}$,''
  Phys.\ Rev.\ D {\bf 89} (2014) 066017
  doi:10.1103/PhysRevD.89.066017
  [arXiv:1312.4542 [hep-th]].
  
\bibitem{Musaev:2015pla}
  E.~T.~Musaev,
  ``Exceptional Field Theory for $E_{6(6)}$ supergravity,''
  TSPU Bulletin {\bf 12} (2014) 198
  [arXiv:1503.08397 [hep-th]].
  
\bibitem{Godazgar:2014nqa}
  H.~Godazgar, M.~Godazgar, O.~Hohm, H.~Nicolai and H.~Samtleben,
  `Supersymmetric E$_{7(7)}$ Exceptional Field Theory,''
  JHEP {\bf 1409} (2014) 044
  doi:10.1007/JHEP09(2014)044
  [arXiv:1406.3235 [hep-th]].
  
\bibitem{Hohm:2014fxa}
  O.~Hohm and H.~Samtleben,
  ``Exceptional field theory. III. E$_{8(8)}$,''
  Phys.\ Rev.\ D {\bf 90} (2014) 066002
  doi:10.1103/PhysRevD.90.066002
  [arXiv:1406.3348 [hep-th]].
  
\bibitem{Rosabal:2014rga}
  J.~A.~Rosabal,
  ``On the exceptional generalised Lie derivative for $d\geq7$,''
  JHEP {\bf 1509} (2015) 153
  doi:10.1007/JHEP09(2015)153
  [arXiv:1410.8148 [hep-th]].
  
\bibitem{Blair:2014zba}
  C.~D.~A.~Blair and E.~Malek,
  ``Geometry and fluxes of SL(5) exceptional field theory,''
  JHEP {\bf 1503} (2015) 144
  doi:10.1007/JHEP03(2015)144
  [arXiv:1412.0635 [hep-th]].



\bibitem{Berman:2014hna}
  D.~S.~Berman and F.~J.~Rudolph,
  ``Strings, Branes and the Self-dual Solutions of Exceptional Field Theory,''
  JHEP {\bf 1505} (2015) 130
  doi:10.1007/JHEP05(2015)130
  [arXiv:1412.2768 [hep-th]].
  
\bibitem{Abzalov:2015ega}
  A.~Abzalov, I.~Bakhmatov and E.~T.~Musaev,
  ``Exceptional field theory: $SO(5,5)$,''
  JHEP {\bf 1506} (2015) 088
  doi:10.1007/JHEP06(2015)088
  [arXiv:1504.01523 [hep-th]].

\bibitem{Berman:2011jh}
  D.~S.~Berman, H.~Godazgar, M.~J.~Perry and P.~West,
  JHEP {\bf 1202} (2012) 108
  doi:10.1007/JHEP02(2012)108
  [arXiv:1111.0459 [hep-th]].
  
\bibitem{Cederwall:2015ica}
  M.~Cederwall and J.~A.~Rosabal,
  ``E$_{8}$ geometry,''
  JHEP {\bf 1507} (2015) 007
  doi:10.1007/JHEP07(2015)007
  [arXiv:1504.04843 [hep-th]].

\bibitem{Baguet:2015xha}
  A.~Baguet, O.~Hohm and H.~Samtleben,
  ``E$_{6(6)}$ Exceptional Field Theory: Review and Embedding of Type IIB,''
  PoS CORFU {\bf 2014} (2015) 133
  [arXiv:1506.01065 [hep-th]].

\bibitem{Tumanov:2015iea}
  A.~G.~Tumanov and P.~West,
  ``E11 and exceptional field theory,''
  arXiv:1507.08912 [hep-th].
  
\bibitem{Bossard:2015foa}
  G.~Bossard and A.~Kleinschmidt,
  ``Loops in exceptional field theory,''
  arXiv:1510.07859 [hep-th].

\bibitem{Hohm:2012gk} 
  O.~Hohm and B.~Zwiebach,
  ``Large Gauge Transformations in Double Field Theory,''
  JHEP {\bf 1302}, 075 (2013)
  [arXiv:1207.4198 [hep-th]].
  
   
\bibitem{Park:2013mpa} 
  J.~-H.~Park,
  ``Comments on double field theory and diffeomorphisms,''
  JHEP {\bf 1306}, 098 (2013)
  [arXiv:1304.5946 [hep-th]].  
  
\bibitem{Berman:2014jba} 
  D.~S.~Berman, M.~Cederwall and M.~J.~Perry,
  ``Global aspects of double geometry,''
  arXiv:1401.1311 [hep-th].
 

\bibitem{Papadopoulos:2014mxa}
  G.~Papadopoulos,
  ``Seeking the balance: Patching double and exceptional field theories,''
  JHEP {\bf 1410} (2014) 089
  doi:10.1007/JHEP10(2014)089
  [arXiv:1402.2586 [hep-th]].
  
\bibitem{Cederwall:2014kxa} 
  M.~Cederwall,
  ``The geometry behind double geometry,''
  arXiv:1402.2513 [hep-th].
  


  
\bibitem{Papadopoulos:2014ifa}
  G.~Papadopoulos,
  ``C-spaces, generalized geometry and double field theory,''
  JHEP {\bf 1509} (2015) 029
  doi:10.1007/JHEP09(2015)029
  [arXiv:1412.1146 [hep-th]].
  
\bibitem{Rey:2015mba}
  S.~J.~Rey and Y.~Sakatani,
  ``Finite Transformations in Doubled and Exceptional Space,''
  arXiv:1510.06735 [hep-th].






  
 

 
   
\end{thebibliography}
\end{document}